\documentclass[%
 aapm,
 mph,%
 amsmath,amssymb,longbibliography,
 reprint,%
]{revtex4-1}

\usepackage{graphicx}
\usepackage{dcolumn}
\usepackage{bm}
\usepackage{amsmath,amssymb}

\usepackage{hyperref}
\usepackage[]{natbib}

\usepackage{tikz}
\usepackage{lipsum}

\begin{document}

\title{Planck scale displacement of a connected mirror pair: an energy measurement in a gravitational field}


\author{Frank V. Kowalski}%
\email{fkowalsk@mines.edu}

\affiliation{Physics Department, Colorado
School of Mines, Golden CO. 80401 U.S.A.}

\date{\today}

\begin{abstract}
Reflection of a particle from a mirror pair that is fixed to a platform which falls freely in a gravitational field is shown to infer displacement of the center of mass of this platform of order the Planck length. This displacement and its relationship to the wave and particle nature of reflection in such an energy measurement is elucidated using harmonic motion of the mirror pair.
\end{abstract}

\maketitle

\section{\label{sec:intro}Introduction}

The Planck length, $L_{P}=\sqrt{\hbar G/c^{3}}$, where $\hbar$ is the reduced Planck constant, $G$ is the gravitational constant, and $c$ is the speed of light, is associated with the scale at which concepts of a metric structure on space become meaningless \cite{garay,hossenfelder,hossenfelder2}. This has stimulated proposals to test the limits on the smoothness of space-time using the displacement of the center of mass, c.m., of a dielectric slab \cite{bekenstein} and to test physics at the Planck scale by probing the canonical commutation relation of the c.m. motion of a mechanical oscillator with that of a Planck mass object \cite{pikovski}. Justification for the use of the c.m. of an object to probe effects at the Planck length rather than its microscopic constituents is given by Bekenstein \cite{bekenstein}. One argument is that the Compton wavelength associated with the c.m. of the slab is smaller than $L_{P}$ whereas that is not the case for the constituents of the mirror pair. 

However, determination of such small mirror displacements is constrained by quantum behavior of the mirror-meter system.  This has been studied extensively \cite{braginsky,aspelmeyer} resulting in a predicted standard quantum limit, SQL, on the accuracy with which a displacement measurement can be made, $\delta X_{SQL}=\sqrt{\hbar \Theta/M}$, where $M$ is the mirror mass and $\Theta$ is the time interval between sequential optical pulses, each of which consists of multiple photons, that interact with the mirror \cite{danilishin}. Two pulses sequentially reflecting from the mirror are used to determine the SQL: measurement of the mirror's position with the first pulse introduces an uncertainty in the momentum of the mirror that, during the time between interaction with the second pulse, causes uncertainty in the position of the mirror and therefore also in the phase of the second pulse in its later reflection. Uncertainty in the number of particles in the pulse constrains the resolution of the phase measurement. Although methods to evade the SQL exist, they do not dramatically exceed the SQL with current technology.  

\begin{figure}
\begin{center}
{\includegraphics[width=6cm]{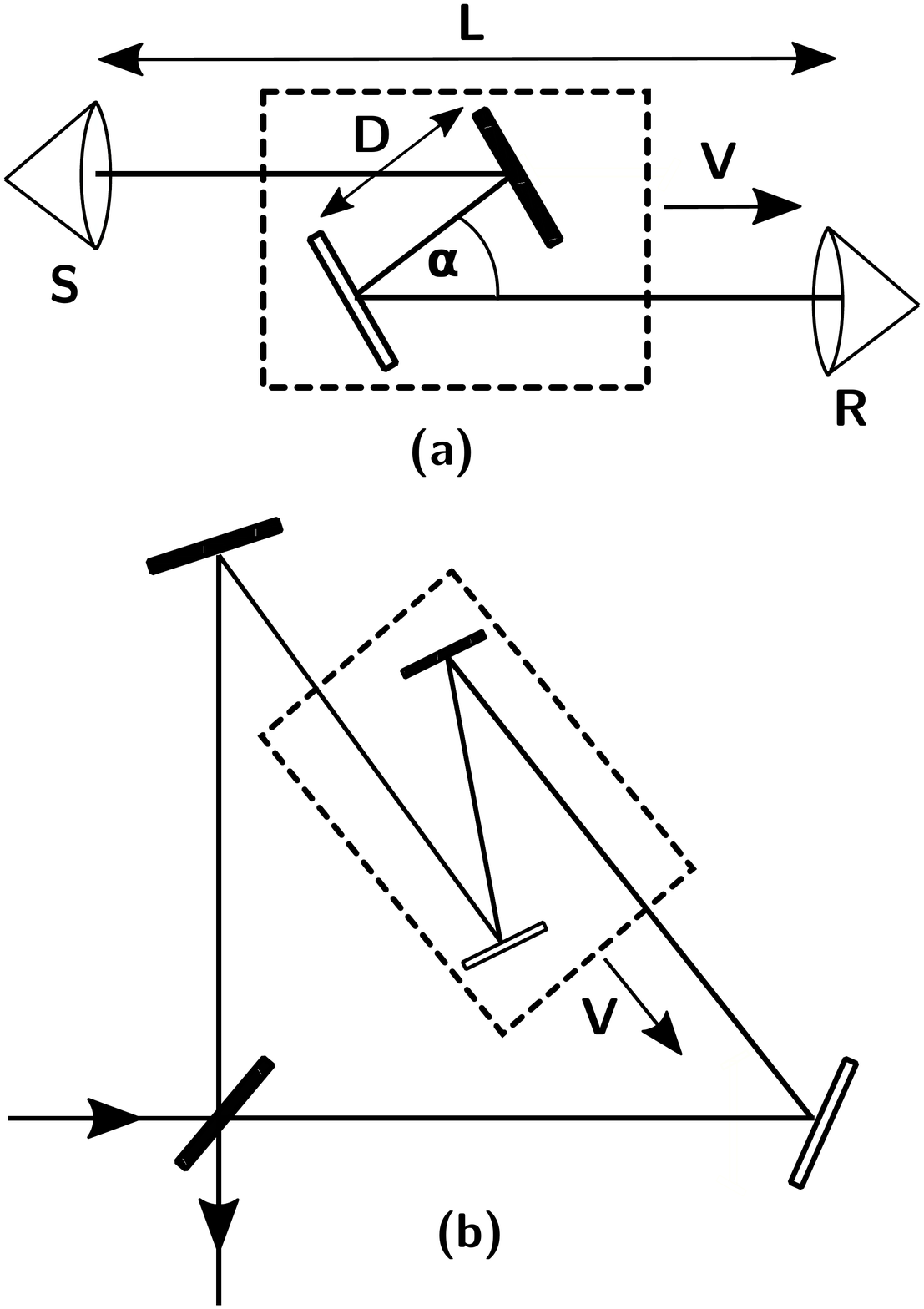}} 
\caption{The dashed rectangle represents the platform on which the mirrors are attached that moves at speed $\textrm{V}$. In (a) the source, S, emits a particle that traverses the mirror pair and is then received at R. Counter-propagating particle states interact with the moving mirror pair in the Sagnac interferometer shown in (b).}
\label{Figure:fig1}
\end{center}
\end{figure}

Fig. \ref{Figure:fig1} illustrates the system considered here: a single particle, generated in an eigenstate of energy $\hbar \omega_{0}$ at S and received at R, reflects from two mirrors that are connected to a platform that is free to move. It is assumed that the path length between S and R does not depend on the small angle of reflection $\alpha$  (the correction term is second order in $\alpha \ll 1$). It is also assumed that no internal energy is generated in the platform due to the reflection and that the momentum of the particle is much less than that of the platform.

The change in phase of the particle wavefunction between traversing a static and moving mirror pair, $\Delta \Phi$, acts as a measure of the speed of the platform. An apparatus that behaves in this manner is often referred to as a speedmeter in the literature and is designed to be a gravitational wave antenna \cite{beyer1,beyer2,chen,khalili,danilishin,freise,danilishin2}. However, the mirrors, analogous to those in fig. \ref{Figure:fig1}(a), move independently rather than being connected to a platform. Additional design features utilize multiple reflections to enhance $\Delta \Phi$ and typically involve a Fabry-Perot arrangement of the mirrors, which in fig. \ref{Figure:fig1} corresponds to $\alpha=0$. 

If the full width half-max, $\textrm{FWHM}$, of the incident particle's probability distribution is much less than the spacing of the mirrors, D, then the interaction is similar to a localized particle generating impulses on the c.m. of the platform, referred to as the C.M. hereafter, during each reflection. This is the regime typically used in deriving the SQL for a speedmeter. On the other hand, if the particle's $\textrm{FWHM}>\textrm{D}$ then the wave-particle interpretation of the reflection is less straightforward. For harmonic motion of the C.M. (described in more detail below) reflection then does not occur suddenly as it would for a localized particle: ``the reflection takes place over a time interval during which the mirror has time to perform several complete oscillations'' \cite{wichman}. The particle's interaction with both mirrors then does not follow a ``classical'' reflection sequence. Therefore, a treatment of the interaction as a localized particle (or a multi-photon pulse) reflecting from the two mirrors is of limited validity. The interaction in a thorough treatment must be described as a wave phenomenon of the system (solving the many-body Schr\"odinger equation) in which the particle's substate $\textrm{FWHM}>\textrm{D}$.


Related many-body calculations of a particle reflecting from a mirror \cite{kowalski1}, a particle interacting with rigid mirrors arranged in a Fabry-Perot configuration \cite{kowalskiwell}, and of two particles reflecting from a mirror \cite{kowalski2} are found in the literature. If only the particle is measured (the results are then predicted by a marginal distribution) and the coherence length of the C.M. is much less than both that of the particle and that of the particle's wavelength then a one-body solution is useful in understanding the quantum behavior of the two-body system \cite{note1}. The discussion here is focused on such a one-body quantum calculation with the particle's substate $\textrm{FWHM}>\textrm{D}$: that of a quantum particle interacting with two classical potentials (a fixed distance apart and free to move) representing the mirrors.

The delay due to the mirror separation can be enhanced by inserting an optical fiber or a slow light medium between the mirrors \cite{dressel}, neither of which, in principle, needs to be rigidly attached to the platform with the mirror pair. Although random variations in the mirror spacing $\textrm{D}$ limit the sensitivity of the mirror pair method to C.M. displacements, they are mitigated in the Sagnac interferometer shown in fig. \ref{Figure:fig1}(b), particularly if these variations in $\textrm{D}$ occur on a time scale larger than the transit time between the mirrors. In addition, since the reflectivity of each mirror is not the same (nor is the absorption coefficient of a dielectric placed between the mirrors zero), a momentum exchange between a particle beam and the C.M. occurs. Such interactions limit the quantum non-demolition nature of the energy measurements on the C.M. Nevertheless, they do not affect the results described below and are therefore not discussed in detail.

A synopsis of what follows is: first, the particle interaction with the mirror pair in the absence of an external force is described. A related calculation is then done that contrasts this methods sensitivity to detecting C.M. motion for the oscillating mirror pair with that for a single oscillating mirror. Next, the interaction of a photon with the mirror pair falling freely in a uniform gravitational field is shown to experimentally infer displacement of the C.M. of order $L_{P}$. A calculation of this effect using a classical electromagnetic field yields a displacement that is inconsistent with the quantum result.

\section{\label{sec:results}Results}

\subsection{\label{sec:const} Mirrors moving at constant speed}

The non-relativistic one-body quantum calculations described below utilize the retarded time method for determining phase shifts \cite{kowalskikowalski}. The particle's phase difference, $\delta \phi$, between the fixed source and receiver is determined by the transit time of a wavecrest between S and R, $\delta t$, where $\delta \phi=\omega_{0} \delta t$. For a particle of mass $m$ emitted from this source with speed $v$, $\delta \phi=m v (L+2\textrm{D})/\hbar+2\textrm{D}m\textrm{V}/\hbar$. For a photon $\delta \phi\approx k (L+2\textrm{D})+2\textrm{D}k\textrm{V}/c$ to first order in $\textrm{V}/c$, where $k$ is the wavevector of the photon. The phase difference between the counterpropagating waves at the output port of the Sagnac interferometer shown in fig. \ref{Figure:fig1}(b) is $ \phi_{Sagnac}=4\textrm{D}m\textrm{V}/\hbar$ or $ \phi_{Sagnac}=4\textrm{D} k \textrm{V}/c$ for massive and massless particles, respectively.

Although this phase difference is a function of $\textrm{V}$, the velocity of the C.M. does not change after the interaction. Rather, the C.M. is displaced due to the delay in the photon traversing the mirror pair, given by \cite{padgett}
\begin{equation}
\delta X = 2\textrm{D} \frac{\hbar \omega_{0}}{M c^{2}},
\label{eq:displacement}
\end{equation}
where $M$ is the mass associated with the C.M.

However, absorption in the mirrors introduces a momentum kick that also displaces the C.M. For $N$ sequential photons traversing the mirrors that have a loss due to scattering and absorption of $\epsilon$ (current optical technology allows one photon to be scattered or absorbed for a million that are reflected or $\epsilon=10^{-6}$), $N\epsilon$ photons generate a recoil that takes a time $\textrm{D}/\epsilon c$ for the C.M. to move the same distance as $N \delta X$. Measurement of the C.M. position for times shorter than this, with an accuracy smaller than the $\textrm{FWHM}$ of the C.M. state, resolves any ambiguity in the cause (absorption or delay) of the displacement. Such absorption does not affect the results described below.

\subsection{\label{sec:osc} Oscillating rigid mirrors}

Next let an external force act on the mirrors shown in fig. \ref{Figure:fig1}(a), generating a C.M. displacement  $x=x_{0} \cos[\Omega t]$. To simplify the calculations only the interaction with photons is considered. Nevertheless, measurements for a neutron traversing an oscillating slab (which generates a delay similar to that of the mirror pair) have been performed \cite{frank,frank2,frank3} while calculations of the interaction of massive particles with accelerating mirrors have been conducted \cite{kowalski3,kowalskikowalski}.

\begin{figure}
\begin{center}
{\includegraphics[width=8cm]{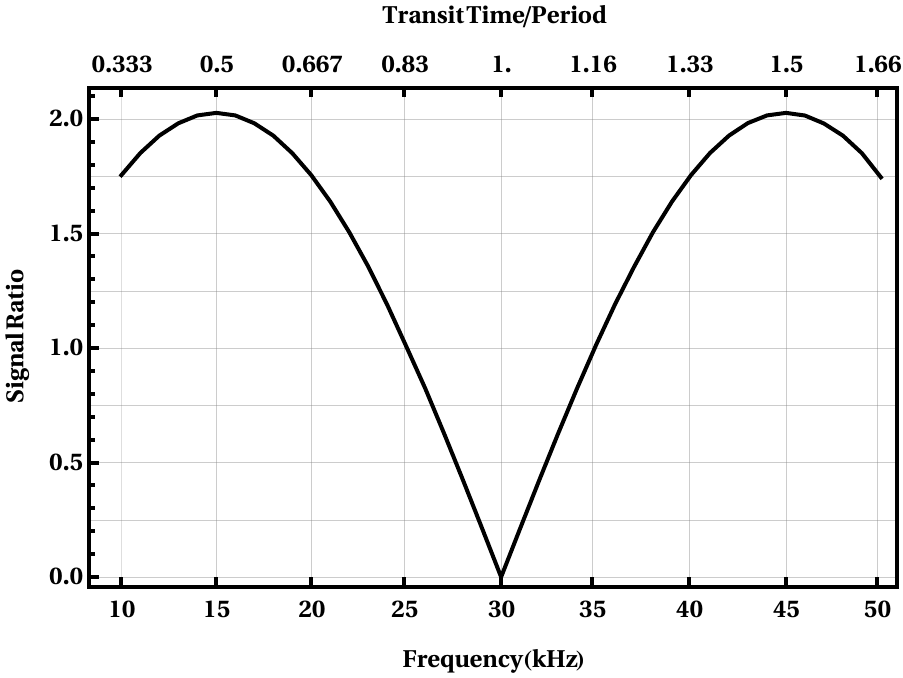}} 
\caption{The ``Signal Ratio'' is the ratio of maximum phase variation of a photon traversing the oscillating mirror pair to that for retro-reflection from one oscillating mirror (with the same $x_{0}$ and $\Omega$). This is plotted vs. either the oscillation frequency (on the lower horizontal frame axis) or the ratio of the transit time of a wavecrest between the mirror pair to the period of oscillation of the mirror pair (on the upper horizontal frame axis).}
\label{Figure:fig2}
\end{center}
\end{figure}

Numerical solutions, as described in the appendix, indicate that the phase of the transmitted wave oscillates sinusoidally as it does in retro-reflecting from a single oscillating mirror. However, for fixed $x_{0}$ the magnitude of this phase oscillation varies with both the period of oscillation, $T=2 \pi/\Omega$, and the transit time between the mirrors, $\tau=\textrm{D}/c$. This is illustrated in fig. \ref{Figure:fig2} for $\textrm{D}=10^{4}$ m and $x_{0}=3 \times 10^{-7}$ m.

The peaks in this graph occur when, for motion of a particular wavecrest, the mirrors appear to move antisymmetrically due to the transit delay. The phase shifts from each mirror then add, thereby generating a phase shift twice that from one mirror. The minimum in this graph indicates symmetric motion of the mirrors for a given wavecrest: the time varying phase introduced by the first mirror is essentially canceled by reflection from the second mirror. At these peaks, for the Sagnac interferometer, the sensitivity to oscillatory motion of the C.M. is four times that of a single oscillating mirror in a Michelson interferometer.

As a heuristic example of a meter in which harmonic motion of the C.M. is manifest let a single optical photon travel between S and R as shown in fig. \ref{Figure:fig1}(a) with the mirror pair oscillating. The output photon state is then analyzed with a grating spectrometer. The photon diffracts at angles associated with the (optical) carrier frequency and with sidebands that are due to the phase modulation (adding small angles to the carrier's angle of diffraction corresponding to multiples of $\Omega$). The probability of measuring a photon at spectrometer angles that are multiples of $\Omega$ decreases as $x_{0}$ decreases. The coherence length of the initial photon state, the grating groove spacing, the width of the spectrometer's slits, and the area of the grating that is illuminated determine if these sidebands are resolved. A measurement of a photon in a sideband infers the existence of C.M. oscillations.

In such a case the C.M. must have experienced a correlated recoil to conserve energy and momentum. Therefore, a two-body (particle and C.M.) quantum state must have both frequency modulated particle and C.M. substates. In this spectrometer measurement the particle nature is manifest in the random appearance of photons at discreet angles and the commensurate recoil of the C.M.  However, the reflection must occur over multiple oscillation cycles of the C.M. rather than instantaneously as with a classical particle to produce such a result.

If, on the other hand, the particle is in a Gaussian substate with $\textrm{FWHM}\ll \textrm{D}$ then the particle's energy spectrum is no longer discrete but rather is determined by the instant at which the peak of the particle substate reflected from each mirror in its oscillation cycle. This yields a result similar to that of a classical interaction: the particle that reflected from each mirror at a retarded time leads to a corresponding recoil of the C.M.  The spectra of sequentially delayed such particle pulses varies essentially continuously. In this manner the wave nature of the interaction imitates that of the classical reflection.


\subsection{\label{sec:grav} Mirrors falling freely in a gravitational field}

Consider a photon with $\textrm{FWHM}\gg \textrm{D}$ reflecting from the mirror pair shown in fig. \ref{Figure:fig1}(a) while all bodies are under the influence of a uniform external gravitational field. The source and receiver are fixed in this gravitational field while the mirror pair falls freely with acceleration $g$ along the direction of $\textrm{V}$ shown in the figure. The Equivalence principle is applied to determine the effect of the gravitational force on this system: S and R accelerate while the photon and mirrors are in an inertial frame. 


The delay of a photon wavecrest in traversing the mirror pair in the inertial frame where the mirror pair is at rest is $2\textrm{D}/c$. The additional wavecrest delay due to traversing the distance between the source and receiver, $L/c$, results in a difference between the photon's energy at the source and receiver of $\hbar \Delta \omega_{EP}=\hbar\omega_{0} g(2\textrm{D}+L)/c^{2}$. This is consequence of the receiver moving at a speed different from that of the source due to the acceleration of the receiver during the transit delay.

Since no external force acts on the photon and mirrors in the inertial frame, the c.m. motion of the particle-mirrors system is undisturbed. The delay of a photon in traversing the mirrors then requires that the C.M. is displaced by $\delta X$ as given in eqn. \ref{eq:displacement}. This displacement is essentially the same as that observed in the frame of the accelerating source and receiver and therefore it is also the same displacement of the freely falling mirrors observed by S and R fixed in the gravitational field.

In the gravitational setting this displacement of the mirror pair results in a loss of potential energy in the gravitation field of $M g \delta X$. For energy to be conserved the photon's energy must increase by a commensurate amount $\hbar \Delta \omega_{displaced} = M g \delta X = \hbar \omega_{0} 2 g\textrm{D}/c^{2}$. In this gravitational setting there is an additional blue shift of the photon due to its change in position $L$ in the gravitational field of $\Delta E_{L}= \hbar \omega_{0} gL/c^{2}$. The net change in photon energy is then the same as that derived from the Equivalence principle above, $\Delta E_{L}+\hbar \Delta \omega_{diplaced}=\hbar \Delta \omega_{EP}$.

An example of such a measurement is found in reference \cite{kowalski0}. The mirror pair was fixed to a platform that moved on an air bearing whose direction of motion was slanted in the gravitational field. This photon-mirrors system was an intracavity component of a He-Ne ring laser which then amplified the time varying phase shift (calculated above) of the counter-propagating waves. The laser was first aligned to operate above threshold with all its components (mirrors and discharge tube) rigidly fixed to a frame. Measurements of the frequency shift were then made after the mirror pair was released to fall freely in the gravitational field, along the slanted angle. The counter-propagating output laser beams were each $\approx 100~\mu$W and the maximum acceleration of the mirrors $\approx 0.1~\textrm{m/s}^{2}$. The apparatus used $M\approx 1$ kg, $\textrm{D}\approx 0.1$ m. The displacement of the C.M. inferred by this measurement of the photon energy is $3.6 \times 10^{-36}$ m while $L_{p}=1.6 \times 10^{-35}$ m. In principle this measurement could have been done with the apparatus shown in fig. \ref{Figure:fig1}(a) by replacing the receiver with the spectrometer apparatus that was described above for the oscillating mirror pair.

The classical interpretation of the interaction between an electromagnetic, E.M., wave and this connected mirror pair is similar to the above quantum result in that it predicts the same frequency shift. However, it differs in improperly accounting for conservation of energy and momentum, both of which are related to the amplitudes of the fields rather than the frequency of the wave. Inference of C.M. displacement using this classical result is next shown to be invalid. 

Although the E.M. wave transports energy classically, it is in quantum theory that this is associated with the transport of mass from S to R. In addition, a comparison between classical and quantum predictions cannot involve a single photon but requires a stream of photons.

To illustrate the discrepancy between the classical and quantum models consider an apparatus similar to that used in the He-Ne ring laser described above, except without there being an external gravitational force. Let a continuous E.M. wave be emitted and received in fig. \ref{Figure:fig1}(a) while S, R, and the mirrors are rigidly connected to a frame. As the E.M. wave continues to be emitted and absorbed, the mirror pair platform is then uncoupled from the frame with $V=0$ and is free to move. Classically, the energy flux received is the same as that emitted. In addition, the radiation pressure on the frame due to emission at S cancels that due to absorption at R. Similarly, the radiation pressure on the platform from the photons reflecting from one mirror negates that from reflection on the other, leaving the C.M. undisturbed.

However, in the quantum interpretation the energy transferred from S to R by the stream of photons results in mass (in the form of the photon energy) leaving S and accumulating at R. The center of masses of the frame and platform are displaced in order for the center of mass of this isolated system to remain unchanged. The stream of photons therefore progressively displaces the frame in a direction opposite to that of the platform. 

In both of these quantum and classical calculations the EM and photon coherence lengths are much greater than the mirror separation. The reflection, unlike that of a point particle, does not generate sequential impulses on the mirror pair platform. Rather, it is similar to reflection from the oscillating mirror pair where the interaction takes place over multiple cycles of oscillation. One might then argue that the displacement is due to the number of photons between the mirror pair rather than from a single photon. In the He-Ne laser experiment the average number of photons interacting with the mirror pair at any given time was $\approx 10^{7}$ photons.  The experiment could be revised using a passive Sagnac interferometer in which one photon on average (using a beam of $c/\textrm{D}$ photons per second) is between the mirrors in the interferometer to generate a displacement of order $L_{p}$.

The displacement of the C.M. is a consequence of the following fundamental principles (that are presumably valid at the Planck scale): the energy-mass relationship and symmetry of the physical laws with respect to boosts. Since this displacement is progressive, a beam of particles traversing the mirror pair will, in principle, eventually result in a mesoscopic C.M. displacement (where there is little doubt about the validity of these principles). The above energy measurement of the photon then indirectly determines the C.M. displacement of order $L_{p}$ while mitigating the effects of back-action in a sequence of measurements.

\section{\label{sec:methods}Appendix}

The oscillating mirror pair is used to illustrate the phase calculations. The transit time, $t_{1}$, for a wavecrest to travel from S to the first mirror is given by $d_{0}+x_{0} \cos[\Omega (t+t_{1})]=c t_{1}$, where $d_{0}$ is the distance from S to the position about which the first mirror oscillates with amplitude $x_{0}$. The transit time for the wavecrest to travel from the first to the second mirror, $t_{2}$, is given by $d_{0}-D+x_{0} \cos[\Omega (t+t_{1}+t_{2})]=c t_{2}$. The distance the crest has to travel from the second mirror to R is given by $L_{final}=L-(d_{0}-D+x_{0} \cos[\Omega (t+t_{1}+t_{2})])$. The transit time for the wavecrest to travel from the second mirror to R is then $t_{3}=L_{final}/c$. These equations are solved numerically to determine the difference in phase between S and R given by $\delta \phi=\omega_{0} (t_{1}+t_{2}+t_{3})$ with the results presented in fig. \ref{Figure:fig2}.


\begin{acknowledgments}
I wish to acknowledge discussions long ago with M. P. Haugen on the interaction of light with accelerating dielectrics. 
\end{acknowledgments}

\nocite{*}
\bibliography{aapmsamp}

\begin{thebibliography}{27}
\bibitem [{not()}]{note1}%
  \BibitemOpen
  \href@noop {} {}\bibinfo {note} {In a two-body quantum treatment the C.M.
  position is uncertain, characterized by its coherence length which is assumed
  to be small enough to have a negligible effect on measurements of only the
  particle. However, this is not the case as the coherence length of the C.M.
  becomes comparable to the wavelength of the incident particle. A related case
  is found in two-body interference between the incident and reflection states
  of a particle and mirror \cite{kowalski1}.}\BibitemShut {Stop}%
\end{thebibliography}

\end{document}